# A Comprehensive Toolbox to Facilitate Quantitative Decision Science in Drug Development: A web-based R shiny application *GOahead*

Bo Wei[1], Weibin Zhong[2], Rong Liu*[2], Alan Wu[3], Alan Chiang[2], Michael Branson[4], Nanxiang Ge[5]

[1]University of Michigan, Ann Arbor, MI 48109, [2]Bristol-Myers Squibb, Berkeley Heights, NJ 07922, [3]Beigene, Ridgefield Park, NJ 07660   [4]UCB Pharmaceuticals, Brussels, Switzerland, [5]Roivant, New York, NY 10036

*corresponding author
Email: rong.liu1@bms.com
Address: 300 Connell Drive
         Berkeley Heights, NJ 07932, USA

## Abstract

Decision making is critical at each stage of drug development and making informed and transparent Go/No-Go decision requires a "sound" quantitative decision framework. We designed and implemented *GOahead*, a comprehensive web-based tool to improve how statisticians and collaborators could prospectively plan and implement the selected Go/No-Go decision approach in real time. In the paper, we conduted comprehensive overview of dual-criterion and confidence interval-based approaches to enable quantitative decision-making. ilillustrative examples are demonstrated for single and two arms designs in both Bayesian and frequentist frameworks, multiple arms design with MCP-MOD is also demonstrated. *GOahead* can be found on shinyapps.io server (*https://goaheadtool.shinyapps.io/GOaheadv10/*)

## Keywords

## 1. Introduction

Drug development naturally has several phases and each phase requires the trial sponsor to address key questions in relation to patient safety and treatment efficacy. The purpose in the early phases are to de-risk the late stage (phase 3) investment as these are usually very time intensive, requires many more patients and a significant financial investment. According to [1], between 2009 and 2018 the estimated mean Research and Development investment required to bring a new drug to market was $1.34 billion (95% CI, $1.04 billion - $1.64 billion), with an average time of more than 10 years. Amidst the huge cost and many years involved in developing a potentially new treatment, there is approximately a 15% probability of success for any new treatment first entering the clinical phase to be successfully registered: averaging across all therapeutic areas between 2003-2011, oncology drugs have the lowest success rate of 7% [2].

The low success rate may be attributed to many reasons and one of the reasons is the high attrition rate of compounds in early development [3]. Thus, emphasis in the early phases of drug development should be to ensure that trials are:

a. Prospectively designed with transparent, reasonable Go/No-Go criteria
b. Clinical trials sized adequately to address these criteria supported by sound statistical principles
c. Analytic software available to calculate the level of evidence supporting the Go/No-Go criteria for key drug development decisions.

In following (a)-(c), it helps inform decision makers whether to progress to the next stage of development or to perhaps reallocate effort to other assets with higher anticipated probability of success.

Implementation of the quantitative decision approach requires close collaboration of clinical team members including clinicians, statistician and colleagues from other functions, to have a clear understanding of what the product target product profile (TPP) (the specification that defines the desired efficacy, safety and other characteristics of the drug product) [4]. Several quantitative decision-making approaches have been proposed and most of these approaches rely on a clearly defined set of rules for a threshold of treatment effect ($\Delta$) and its probability distribution $P(\Delta)$ ([5], [6]). For instance, the threshold $\Delta$ may be the treatment effect of a drug relative to a placebo or an active comparator in a phase II trial and the defined set of rules are referred to as the decision criterion.

Quantitative decision-making approaches from authors such as [7] and [8] focused on both the statistical significance and clinical relevance of $\Delta$ in making a Go/No-Go decision in what is termed the dual-criterion design, and can be established in both a Bayesian and frequentist framework. Other approaches focus on the lower and upper cases of TPP to define a "multilevel" threshold ([4], [9]). We propose a unified framework for approaches that encompasses the existing Bayesian and frequentist methods and provide an extension to include various data types and clinical trial design settings.

There is no available package or procedure in commonly used statistical software such as R or SAS to readily implement the Go/No-Go framework, especially when the study has Go/No-Go on multiple correlated endpoints and with interim analysis. Therefore, we developed and validated through independent programming a user-friendly tool (referred to as *GOahead*) providing flexible quantitative decision-making solutions that is designed and implemented in a R shiny environment. *GOahead* has improved how statisticians and project teams at Bristol Myers-Squibb interact as trials are prospectively planned and Go/No-Go decisions can be calculated and communicated with clinical development team in real time. *GOahead* implements all the quantitative decision-making approaches mentioned above and can handle a total of 18 different decision criteria constellations. The rest of the paper is organized as follows: section 2 gives a comprehensive description of specific methods; section 3 reports the results and section 4 provides the discussion and conclusion.

**2. Methods**

2.1. Background

Clinical drug development is becoming more complex and the pharmaceutical industry is increasingly replying on the use of more quantitative approaches to help meet the challenges. In this section, we look at a brief background of the various quantitative decision-making approaches that *GOahead* implements and how each approach makes its Go/No-Go decisions and on how one approach compares to another. All the current approaches rely on the use of a single primary endpoint to define the Go/No-Go criteria.

Roychoudhury et al. proposed the dual-criterion design [8]. The dual criterion combines statistical significance with clinical relevance and is centered around two key values, the null value (NV; value under a specified null hypothesis) and the decision value (DV: a clinically minimum effect estimate needed for trial success) as benchmarks. In the frequentist aspect, statistical significance here represents that a one-sided p-value is less than the significance level $\alpha$ or a one-sided $100(1-\alpha)\%$ confidence interval excludes the NV. In the Bayesian aspect, it represents that the posterior probability of the effect that is greater than the NV is greater than a prespecified confidence $1-\alpha$. With the minimum clinical relevance, the estimated effect or the posterior median effect must reach at least the DV for the frequentist approach or the Bayesian approach, respectively. Let $LCB$ and $UCB$ be the lower confidence bound and the upper confidence bound, respectively. An example of dual criterion rule by using the frequentist approach may be:

$$1. One-sided\ 95\%\ LCB(\Delta) \geq NV;$$
$$2. \Delta_{obs} \geq DV,$$

Another example of dual criterion rule by using the Bayesian approach may be:

$$1. Pr(\Delta \geq NV|data) \geq 95\%;$$
$$2. Pr(\Delta \geq DV|data) \geq 50\%.$$

From the examples, $\alpha$ is set at 0.05, and $\Delta$ could "quantify the primary efficacy of an experimental drug in a single arm design" [8]. For the frequentist approach, a Go decision requires that $\Delta$ must be greater than the NV with significant level $\alpha$ and its observed statistic must be numerically greater than the DV. For the Bayesian approach, a Go decision requires a high probability to achieve statistical significance better than the NV and a moderate probability (50%) to satisfy the DV.

With the dual criterion design, there are three possible decision outcomes: a Go decision, a No-Go decision and an Inconclusive decision. A Go decision is reached with this design if the observed data meets both the statistical significance and clinical relevance. A No-Go decision on the other hand is reached if the observed data fail to meet both the statistical significance and the minimum clinical relevance. If either one of statistical significance or minimum clinical relevance is met, then the decision results in an Inconclusive outcome.

Another frequentist quantitative decision-making approach was initially proposed by [9] and was later adapted by [10]. Such approach to decision-making revolves around two very important values, the target value (TV) and the lower reference value (LRV). According to [10], TV "is set at the desired effect to potentially establish a compound as the treatment of choice". A typical example of a TV may be a market estimate of commercial viability [9]. The LRV on the other hand may be defined as the "smallest clinically meaningful treatment effect for the development of a compound" or the "dignity line". For example, LRV may be the treatment effect of a competitor compound known to have a mild efficacy [9]. With the decision-making approach in [10], the upper and lower cutoff points are obtained for the estimated treatment effect $\hat{\Delta}$ given the sample size, pre-specified false GO risk (FGR) given LRV as null (similar to type I error) and false NOGO risk (FNGR) given TV as alternative (similar to type II error). The approach has three possible decision outcomes just as we saw with [8]: a Go decision, an Inconclusive decision and a No-Go decision. In [9] and [10], the Inconclusive and the No-Go decisions are referred to as Consider and Stop decisions respectively. The Inconclusive decision is anything between the Go and No-Go criteria. The three decisions in [10] can be defined as:

$$GO: One-sided\ (1-\alpha)\%\ LCB(\Delta) \geq LRV;$$

$$NOGO: One-sided\ \beta\%\ LCB(\Delta) < TV;$$

$$Inconclusive: otherwise,$$

where $\alpha$ and $\beta$ are the pre-defined FGR and FNGR, respectively.

Pulksteins et al extends [10] through a Bayesian paradigm [4]. Fig 1 (as seen in [10]) is a visualization of this approach.

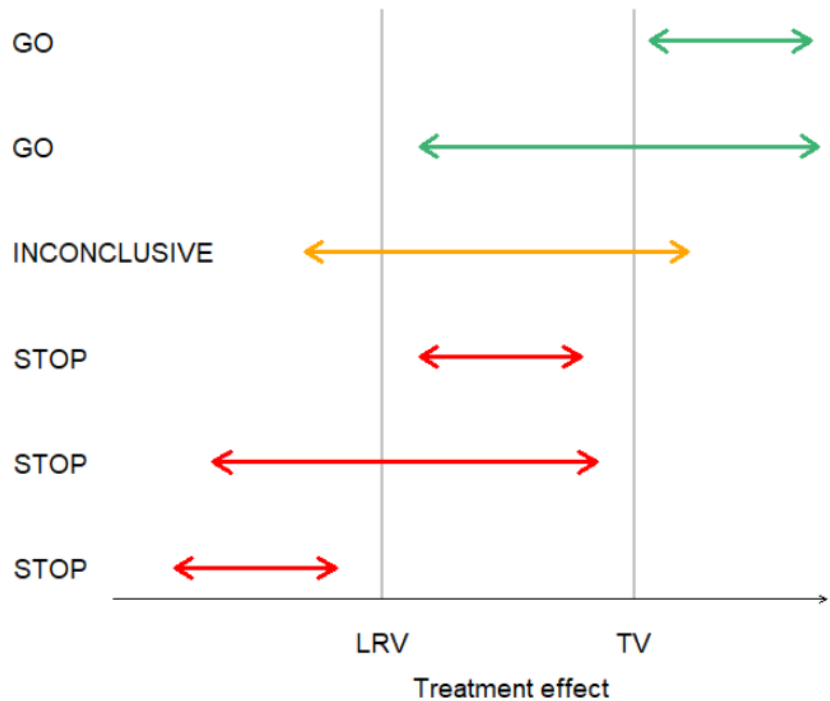

**Fig 1**. Decision framework for Go, Inconclusive and No-Go decisions (LRV: Lower reference value, TV: Target value)

The three decisions in [4] are formally defined as:

$$GO: PCT_{20} > LRV \text{ and } PCT_{90} > TV$$
$$Inconclusive: PCT_{20} \leq LRV \text{ and } PCT_{90} > TV$$
$$No-Go: PCT_{90} \leq TV$$

where $PCT_x$ is the $x^{th}$ percentile of $P(\Delta|data)$. Notice that this approach is equivalent to the Bayesian approach proposed by [8] if we treat TV as DV and LRV as NV.

The NV in [8] and the LRV in [10] both refer to the value that treatment effect needs to be better to continue further development such as the "smallest clinically meaningful treatment effect for the development of a compound" or the "dignity line". The authors ([9], [10]) describes TV as closely related to TPP that is the desired effect to potentially establish a compound as the treatment of choice. In comparison, DV is the minimum treatment effect that needs to be observed in the designed study, which is generally lower than TV [8]. However, in early phase development (Phase I or II studies), TV is most likely not directly related to the TPP unless there is an intermediate endpoint that can be used for registration. Furthermore, when both TV and DV refers to an intermediate endpoint, it is more intuitive to set up a DV that needs to be "observed" as opposed that a TV that may only be supported with, say, 20% confidence to achieve a Go decision.

## 2.2. Description of *GOahead*

*GOahead* is a tool that is used by statisticians/clinicians in quantitative decision science (QDS) planning and analysis of clinical trials to support decision making. It is implemented with an R shiny interface, an R studio application. Four designs implemented in *GOahead* are single arm design, two arms design, multiple arms (as in dose-finding study) design and two correlated endpoints design (for single arm design). The multiple arms design is implemented with the Multiple Comparison and Modeling (MCP-MOD) methodology by [11]. The tool also implements four different design types where a design can be with an interim analysis or without interim analysis.

Users can choose three different data types: binomial, normal and survival, each of which can be implemented in either a frequentist [10] or a Bayesian framework [8] [4].

Table 1 is a summary of the data types, inference frameworks and the design implemented in *GOahead*.

**Table 1**

Summary of data and designs types implemented in *GOahead*

| Design | Data Type | Frequentist | | Baysesian | |
|---|---|---|---|---|---|
| | | Standard analysis | Interim analysis | Standard analysis | Interim analysis |
| Single arm | Binary | Yes | Yes | Yes | Yes |
| | Normal | Yes | Yes | Yes | Yes |
| | Survival | Yes | Yes | Yes | Yes |
| Two arms | Binary | Yes | Yes | Yes | Yes |
| | Normal | Yes | Yes | Yes | Yes |
| | Survival | Yes | Yes | Yes | Yes |
| Multiple arms | Binary | Yes | Yes | No | No |
| | Normal | Yes | Yes | No | No |
| Two correlated endpoints | Binomial | Yes | Yes | Yes | Yes |
| | Normal | Yes | Yes | Yes | Yes |

Based on user specified decision criteria, the *GOahead* tool generates cut off/s for the selected study/design and thereafter computes the probabilities for Go, No-Go and Inconclusive (if relevant) using the generated cut off/s. *GOahead* is a very flexible design and analysis QDS tool in that it allows for various approaches to decision criteria. For instance, some commercial software that implements the frequentist approach [10] and the Bayesian approaches [4] [8] cannot be used to implement a study design with decision criterion

$$Go\ if\ One-sided\ X\%\ LCB(\Delta) \geq NV\ or/and\ One-sided\ Y\%\ LCB(\Delta) \geq DV$$

$$No-Go\ if\ One-sided\ Z\%\ LCB(\Delta) < NV\ or/and\ One-sided\ W\%\ LCB(\Delta) < DV$$

and

$$Go\ if\ Pr(\Delta \geq NV|data) \geq X\%\ or/and\ Pr(\Delta \geq DV|data) \geq Y\%$$

$$No-Go\ if\ Pr(\Delta \geq NV|data) < Z\%\ or/and\ Pr(\Delta \geq DV|data) < W\%$$

respectively.

*GOhead* overcomes this drawback because it is designed to implement several decision rules simultaneously. Table 6 in the supplementary material shows how *GOahead* chooses cut off/s when Go rule has two criteria and No-Go rule also has two criteria as can be seen in the above decision criterion. For a Go rule: *GOahead* allows users to input one or two decision criteria with the confidence of Go (%) for each criterion, choose the relationship between the two Go criteria (with logical operators "and/or") and also indicate the direction of comparison (logical operators: "greater or equal/less"). In a similar manner, users can also input one or two No-Go rules each with its own false No-Go risk (%), decide the relationship between the two No-Go criteria ("and/or") and the direction of comparison ("greater or equal /less"). In totality, the *GOahead* tool implements 18 different quantitative decision-making criteria as shown in Fig 2. The flexibility of greater or equal/less option allows *GOahead* handle both efficacy and safety endpoints.

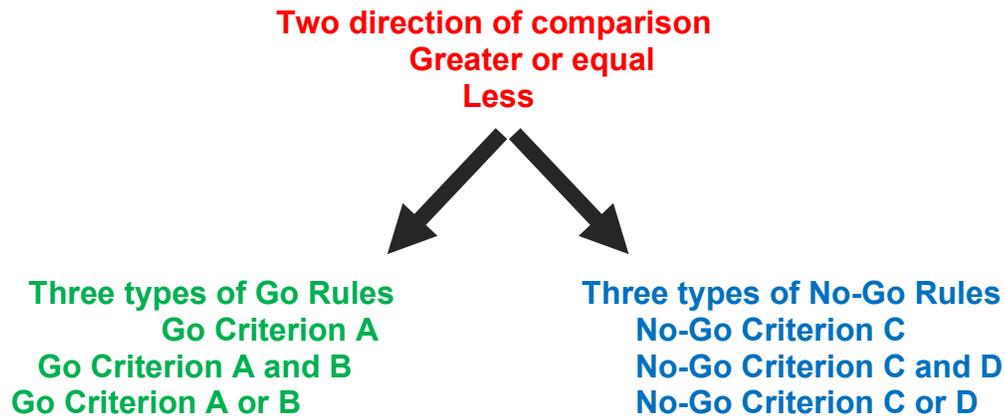

**Fig 2.** Number of decision rules implemented in *GOahead*: 2*3*3 = 18 different rules

2.3. Technical details of each design/module in *GOahead*

The *GOahead* produces three main outputs when in use:

a. cut off/s

Cut offs for some modules are calculated using analytical methods while others are calculated through simulations. Regardless of which method was used to get the cut off/s, the probabilities of Go, No-Go and Inconclusive are always calculated analytically with specific distributional assumptions of the parameter of interest (see Table 2 and Table 3). For a single arm binomial design with a calculated cut off, the tool uses the binomial probability mass function to calculate the probabilities of Go, No-Go and Inconclusive. The exception to this is the multiple arms module where no cut off is calculated and the probabilities of Go, No-Go and Inconclusive are computed through simulations. With the multiple arms module, two keys aspects of the MCP-MOD methodology, MCP step (used to establish POC) and MOD step (used to calculate a target dose) (refer to [11], [12]) were utilized in computing the probabilities of Go, No-Go and Inconclusive zones of the quantitative decision-making process. Table 2 and Table 3 give summaries of the theoretical detail for each design/module. Table 4 also shows whether a cut off for a design is calculated through simulation or via an analytical method.

b. Operating characteristics of the design

With each design in *GOahead,* aside the cut off/s, users are also able to copy or download the operating characteristics (probabilities of Go, No-Go and Inconclusive) of each design specification along with a warning that will let users know if the Go and No-Go criteria overlap. When Go No-Go criteria overlap, *GOahead* allow users to choose which of the two criteria dominates.

c. Graphical representation of designs

For most designs or modules, three different graphs are produced, one showing the cut off/s given the observed parameter of interest (example response rate, mean, difference in means etc.) for Go and No-Go with varying sample sizes, another graph shows the probabilities of Go, No-No and Inconclusive with varying sample sizes, and a third graph shows the probabilities of Go, No-Go and Inconclusive across different values of parameter of interest.

**Table 2**

Theoretical details for each frequentist design/module

| Design | Data Type | Estimator | Frequentist Distribution | Comment |
|---|---|---|---|---|
| Single arm | Binary | $\hat{p} = \dfrac{X}{n}$ | $X \sim Binomial(n,p)$ | exact one-sided confidence interval for cut off |
| | Normal | $\bar{X}$ | $\bar{X} \sim N\left(\mu, \dfrac{\sigma^2}{n}\right)$ | |
| | Survival | $SP$ | $SP \sim \exp(\lambda)$ $X \sim Binomial(n, SP)$ | users can choose between two distribution for this module |
| Two arms | Binary | $\widehat{p_t} - \widehat{p_c}$ | normal approximation of $\widehat{p_t} - \widehat{p_c}$ | |
| | Normal | $\overline{X_t} - \overline{X_c}$ | $\overline{X_t} - \overline{X_c} \sim$ $N(\mu_t - \mu_c, SE)$ | $SE = \sqrt{\left(\dfrac{\sigma_t^2}{n_t}\right) + \left(\dfrac{\sigma_c^2}{n_c}\right)}$ |
| | Survival | $\widehat{HR}$ | $\log(\widehat{HR}) \sim$ $N(\log(HR), SE)$ | $SE = \sqrt{\dfrac{1}{m_1} + \dfrac{1}{m_2}}$ |
| Multiple arms (MCPMOD) | Binary | optimal contrast | multivariate t-test for Proof-of-Concept test | data is generated with rbinom |
| | Normal | | | data is generated with rnorm |
| Two correlated endpoints | Binomial | observed cell counts in 2x2 table | $(n_{00}, n_{01}, n_{10}, n_{11})$ $\sim Multinomial$ $(p_{00}, p_{01}, p_{10}, p_{11})$ | cut off is based on marginal distribution of endpoint |
| | Normal | mean for each endpoint | $Endpoint\ 1:$ $X \sim N(\mu_x, \sigma_x^2)$ $Endpoint\ 2:$ $Y \sim N(\mu_y, \sigma_y^2)$ $(X, Y) \sim N_2\left((\mu_x, \mu_y), \Sigma\right)$ | $\Sigma$ is the variance-covariance matrix |

MCPMOD: multiple comparisons and modeling by Bretz et al., 2009
X = Number of responses; SP = Survival Probability; SE = Standard Error; HR = Hazard Ratio; n = sample size; $n_c$ = sample size in control group; $n_t$ = sample size in treatment group; $m_1$ = number of events in control group; $m_2$ = number of events in treatment group
13

**Table 3**

Theoretical details for each Bayesian design/module

| Design | Data Type | Estimator | Bayesian Distribution | Comment |
|---|---|---|---|---|
| Single arm | Binary | $X \sim Binomial(n, p)$ | $p \sim Beta(\alpha, \beta)$ | |
| | Normal (Known variance) | $Y \sim N(\mu, \sigma^2)$ | $\mu \sim N(\mu_0, \sigma_0^2)$ | |
| | Normal (Unknown variance) | | $\mu \sim N\left(\mu_0, \frac{k_0}{\sigma^2}\right)$ $\frac{1}{\sigma^2} \sim Gamma(\alpha_0, \beta_0)$ | users can also choose the Jeffery's prior with this design |
| | Survival | $X \sim Binomial(n, SP)$ | $SP \sim Beta(\alpha, \beta)$ | SP is the survival probability |
| Two arms | Binary | $X_t \sim Binomial(n_t, p_t)$, $X_c \sim Binomial(n_c, p_c)$ | $p_t \sim Beta(\alpha_t, \beta_t)$ $p_c \sim Beta(\alpha_c, \beta_c)$ | focus is on $p_t - p_c$ |
| | Normal (Known variance) | | $\mu_t \sim N(\mu_{0t}, \sigma_{0t}^2)$ $\mu_c \sim N(\mu_{0c}, \sigma_{0c}^2)$ | $SD = \sqrt{\left(\frac{\sigma_t^2}{n_t}\right) + \left(\frac{\sigma_c^2}{n_c}\right)}$ |
| | | $\bar{X} - \bar{Y} \sim N(\mu_t - \mu_c, SD)$ | $\mu_t \sim N\left(\mu_{0t}, \frac{k_{0t}}{\sigma_t^2}\right)$ | |
| | Normal (Unknown variance) | | $\frac{1}{\sigma_t^2} \sim Gam(\alpha_{0t}, \beta_{0t})$ $\mu_c \sim N\left(\mu_{0c}, \frac{k_{0c}}{\sigma_c^2}\right)$ $\frac{1}{\sigma_c^2} \sim Gam(\alpha_{0c}, \beta_{0c})$ | |
| | Survival | $\log(\widehat{HR}) \sim N(\log(HR), SE)$ | $\log(HR) \sim N(\mu_0, \sigma_0^2)$ | $SE = \sqrt{\frac{1}{m_1} + \frac{1}{m_2}}$ |
| Two correlated endpoints | Binomial Normal | cut off calculation is same as in the single arm for each endpoint and probabilities for Go, No-Go and Inconclusive are calculated using either bivariate Binomial or multivariate normal | | |

X = Number of responses; SP = Survival Probability; SE = Standard Error; HR = Hazard Ratio; n = sample size; $n_c$ = sample size in control group; $n_t$ = sample size in treatment group; $m_1$ = number of events in control group; $m_2$ = number of events in treatment group

**Table 4**

Methods for calculating cut off/s and probabilities for each module

| Design | Data Type | Frequentist | | Bayesian | |
|---|---|---|---|---|---|
| | | Cut off analytical | Probability analytical | Cut off analytical | Probability analytical |
| Single arm | Binary | analytical | analytical | analytical | analytical |
| | Normal (Known variance) | analytical | analytical | analytical | analytical |
| | Normal (Unknown variance) | analytical | analytical | analytical | simulation |
| | Survival | analytical | analytical | analytical | analytical |
| Two arms | Binary | analytical | analytical | simulation | simulation |
| | Normal (Known variance) | analytical | analytical | analytical | simulation |
| | Normal (Unknown variance) | analytical | analytical | simulation | simulation |
| | Survival | analytical | analytical | analytical | simulation |
| Multiple arms | Binary | | simulation | | |
| | Normal | | simulation | | |
| Two correlated endpoints | Binomial | analytical | simulation | analytical | analytical |
| | Normal (Known variance) | analytical | analytical | analytical | analytical |
| | Normal (Unknown variance) | analytical | analytical | simulation | simulation |

## 2.4. Computation environment

The *GOahead* QDS tool can be run on any local computer requiring base R and R studio installed. In addition, the following inbuilt R packages; "cluster, mvtnorm, shiny, lattice, foreach, doParallel" are needed to be installed to run *GOahead*. Additionally, MCPAPP_0.1.0.tar.gz is a user defined package that needs to be installed. Nonetheless, *GOahead* is available on RStudio's shinyapps.io server and can be accessed at [https://goaheadtool.shinyapps.io/GOaheadv10/](https://goaheadtool.shinyapps.io/GOaheadv10/).The *GOahead* app can handle flexible Go No-Go design rules, do efficient computation and produce friendly and transparent output for future analysis. Modules or designs such as two arms Binomial and two arms normal with Bayesian approaches and multiple arms design require extensive simulations. Generally, for the two arms Bayesian designs, a 2000 simulation size will take about 12 -15 seconds to generate desired outputs. Multiple arms design (implemented with MCP-MOD) on the contrary can take about 2-8 minutes for the same 2000 simulation size. Since parallel computing is enabled for the functions behind the *GOahead* app, the computation time can be shorter when more cores are provided. However, this is still one of the areas of improvement we look forward to in our future work, to cut down on the simulation times for some of these modules.

## 2.5. Validation for *GOahead*

Validation for *GOahead* was done in house and each design or module was independently validated by two statisticians. For each module, both statisticians wrote independent code (whether in R or SAS) and based on a hypothesized or a published example, generated outputs with their codes. The outputs generated from the independent code and those generated from *GOahead* were compared. With modules that use exact methods, the goal was to have outputs generated from both *GOahead* and the independently written codes to be the identical. In contrast, with modules that rely on simulations, the goal was to have outputs generated from both *GOahead* and the independent codes to be close, with acceptable Monte Carlo error.

## 3. Results

Example 1: Single Arm Binary Endpoint Frequentist Analysis

As our first illustrative example, we demonstrate a simple quantitative decision-making criterion for Renal Cell Cancer (RCC) Phase IIa single arm dose expansion study evaluating overall response rate [13] using the frequentist method. The planned sample size was 25 and the decision criteria, which has been revised are:

***Decision criteria:***
$$Go: One-sided\ 80\%\ LCB(ORR) \geq 0.2$$
$$No-Go: One-sided\ 10\%\ LCB(ORR) < 0.3$$
$$\textit{Inconclusive: otherwise}$$

We are interested in evaluating the design operating with the specified sample size and the revised criteria if the drug's true ORR rate is 0.15, 0.2, 0.3, or 0.4. Fig 3, Fig 4 and Fig 5 show what users are to input into the *GOahead*:

**Fig 3.** Data type, design and Go criterion for single arm Binomial example

**Fig 4.** No-Go criterion and decision rules for single arm Binomial example

## Parameters

### Interested sample sizes and true values

**Interested sample size**
25

**Interested values**
0.15,0.2,0.3,0.4

**The sample size for OC figure w.r.t different true values**
25

**The true value for OC figure w.r.t sample size**
0.3

Go

**Fig 5.** Parameters and interested/true values for single arm Binomial example

With these input specifications, the *GOahead* generates the following outputs and users have the flexibility to extract all these outputs.

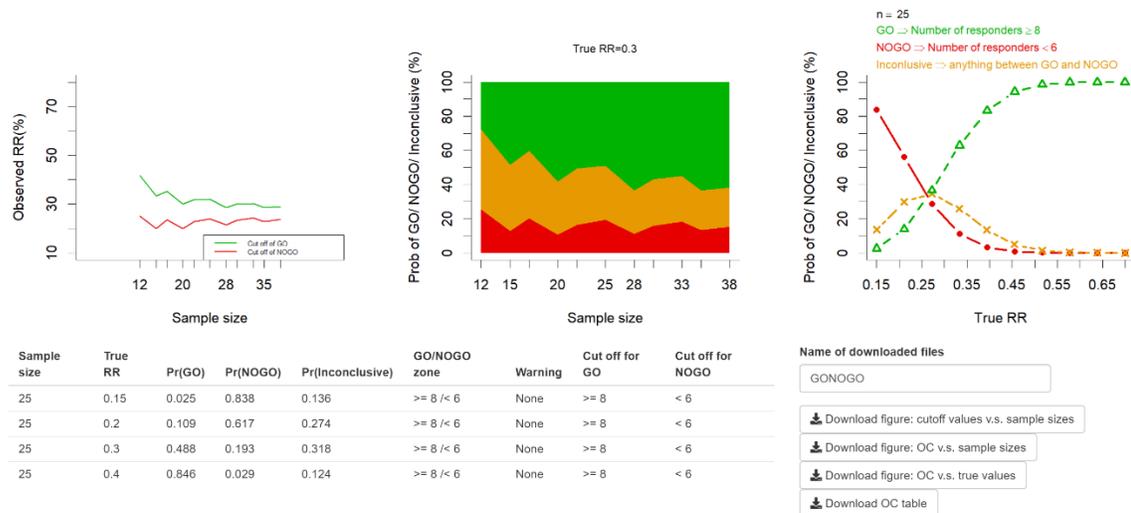

**Fig 6.** Downloadable outputs for single arm Binomial example

Example 2: Two Arm Survival Endpoint Frequentist Analysis

As a second illustration, example 2 is a randomized double-blind Proof-of-Concept trial of an experimental drug in combination with standard of care in patients with non-small-cell lung cancer in [8]. The primary endpoint here was Progression-Free-Survival (PFS). In this example, the authors in [8] assumed 1:1 allocation ratio, and the approximate normality of log hazard ratio. In *GOahead*, we use the hazard ratio in two-arm survival module for calculation. The quantitative decision criteria as we can see with this example is the dual criteria. With the dual criteria, the *GOahead* gives users the flexibility to choose either an "or" or an "and" relationship between the two rules for both the Go and No-Go. The preambles for this example are:

$ln(HR) \sim N(ln(HR), \sigma)$, with $\sigma = \sqrt{(1/m_1) + (1/m_2)}$, where $m_1 = m_2 = \frac{70}{2} = 35$, and we assumed true HR=0.5,0.6,0.7.

***Decision criteria:***

$$\textcolor{green}{Go: One-sided\ 90\%\ UCB(HR) \leq 1}$$
$$\textcolor{green}{and\ One-sided\ 50\%\ UCB(HR) \leq 0.7}$$
$$\textcolor{red}{No-Go: One-sided\ 90\%\ UCB(HR) > 1}$$
$$\textcolor{red}{and\ One-sided\ 50\%\ UCB(HR) > 0.7}$$
$$\textcolor{orange}{Inconclusive: otherwise}$$

Fig 7- Fig 10 give the user inputs and outputs from the *GOahead*

**Fig 7.** Go criterion for hazard ratio in two arms survival example

**Fig 8**. No-Go and decision criteria for hazard ratio in two arms survival example

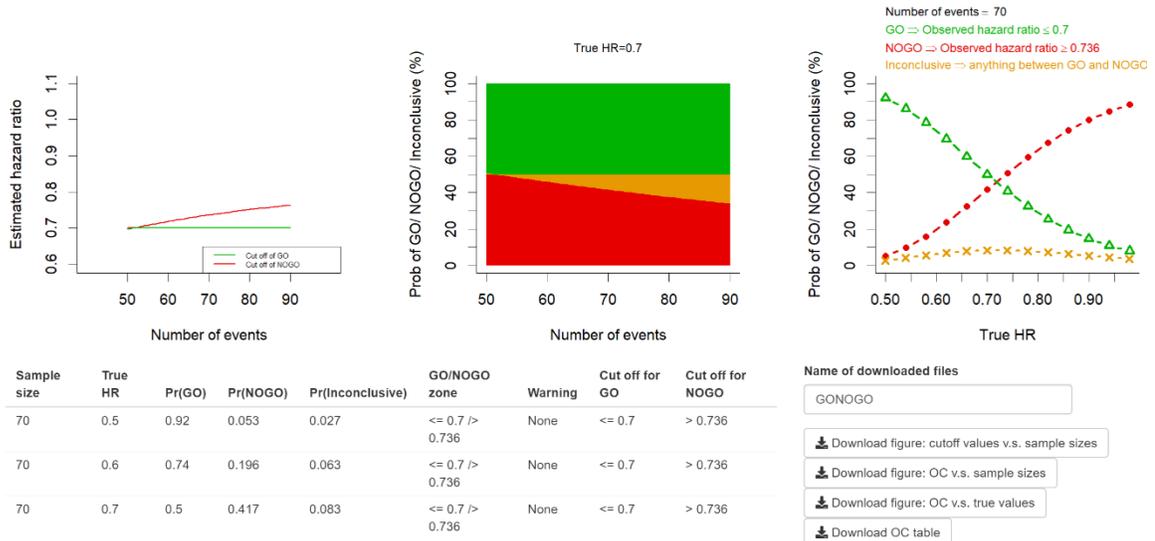

**Fig 9.** Parameters and interested values for hazard ratio in two arms survival example

**Fig 10**. Downloadable outputs for hazard ratio in two arm survival example

Example 3: Single Arm Normal Endpoint, Bayesian Analysis

Finally, we illustrate a hypothetical example on a single arm study with a normal endpoint where inference is done in the Bayesian framework. We assume an unknown mean and a known variance for the population mean and the population variance respectively. The input parameters and the decision criteria needed for this design are: $X \sim N(\mu, \sigma)$; $\mu \sim (0.333, 1)$, $\sigma = 1$; n = 30,45,50; interested true mean = 3.8

***Decision criteria:***

$$Go: \Pr(\mu \leq 4|\bar{X}) \geq 90\% \ or \ \Pr(\mu \leq 2.5|\bar{X}) \geq 50\%$$
$$No - Go: \Pr(\mu \leq 4|\bar{X}) < 20\ \%$$
$$Inconclusive: otherwise$$

Fig 11, Fig 12 and Fig 13 show how the parameters and decision criteria are input in *GOahead* and Fig 14 is what the output is.

**Fig 11**. Go criterion for single arm Bayesian normal example

**Fig 12**. *No-Go and decision criteria for single arm Bayesian normal example*

**Fig 13**. *Prior information and interested values for single arm Bayesian normal example*

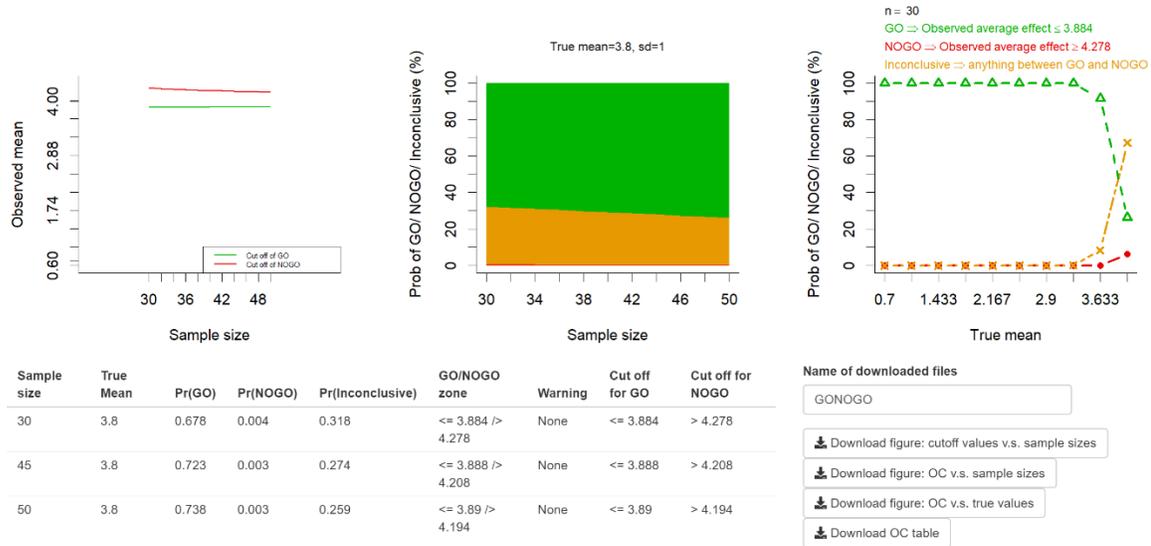

**Fig 14.** Downloadable outputs for single arm Bayesian normal example

Notable among Fig 3, Fig 7 and Fig 11 is the stop criterion for calculation and the random seed for reproducible results. The former is a positive number which is used to control the accuracy of computation for the cut offs for Binomial two arms modules both in the frequentist and Bayesian approaches. It is recommended to make this number as small as possible although the smaller it is, the longer the computation takes.

With Fig 4, Fig 8 and Fig 12, we also notice the Dominated rule section. Here, a user can either select "Go" or "No-Go" as the rule *GOahead* should follow if the Go rule and the No-Go rule are in conflict.

The downloadable outputs in Fig 6, Fig 10 and Fig 14 further demonstrate the simplistic outputs which allow users ease of interpretation. In the supplementary material, we further demonstrate *GOahead* with multiple arms binary endpoints, two arms normal endpoints with Bayesian inference and two arms survival endpoints with interim futility analysis.

## 4. Discussion and Conclusion

Drug development is complex and there is an increase use of more quantitative approaches to facilitate product milestone decisions. In this paper, we extended the existing quantitative decision approaches in clinical development and presented a tool that has been validated through independent programming. It provides a flexible quantitative decision-making tool referred to as *GOahead* and is implemented in a user friendly R shiny environment.

There are a few limitations worth noting in terms of the algorithm *GOahead* tool uses for some of the designs. The algorithm used to implement the two arms binomial design in the frequentist framework is complex. We assumed a normal approximation and the cut off/s were calculated based on this assumption. With interim analysis, *GOahead* relies on simulation study to calculate the probabilities Go/No-Go. Looking ahead, we could use a closed form algorithm for calculating the Go No-Go probabilities in interim analysis.

*GOahead* affords statisticians, clinicians and cross functional teams involved in designing and making transparent and information decision a straightforward and versatile tool with graphics that can support decision making and communication with key stakeholders.

*GOahead* tool was designed to primarily address early and early transitioning to late phase development and exploratory study questions. It was not particularly intended for confirmatory studies where other statistical tools such as FACTS® and EAST® can be readily applied.

Overall, *GOahead* tool covers 4 designs (single arm, two arms, multiple arm and correlated endpoint), with or without interim analysis, and different endpoints such as normal, binary and survival. It also support Muti-arm dose-finding designs. *GOahead* can be used not only for efficacy endpoints, but also for safety endpoints. It allows both frequentist and Bayesian analysis. Undoubtedly, as established in the illustrative examples, *GOahead* can handle multiple quantitative decision-making criteria and aims to address a comprehensive consideration in study design and is not limited to only one form of design criteria based design. *GOahead* has improved how statisticians and project team interact and was proven to be a useful tool to facilitate quantitative decision science at Bristol-Myers Squibb.

**Conflict of interest statement**

No potential conflicts of interest were disclosed by the authors

**Acknowledgements**

Authors thank BMS colleagues Ahrim Youn, Revathi Ananthakrishnan, Tianhua Wang, Yanping Chen, Frank Shen, Angeliki Zarotiadou, Eunice Ampah, Yujie Zhao for their support on the development of GOahead, especially their efforts on validation of GOahead.

**Funding**

This work was supported by Bristol-Myers Squibb

## 5. References

[1] O.J. Wouters, M. McKee, J. Luyten, Estimated Research and Development Investment Needed to Bring a New Medicine to Market, 2009-2018, JAMA 323(9) (2020) 844-853. https://doi.org/10.1001/jama.2020.1166.

[2] M. Hay, D.W. Thomas, J.L. Craighead, C. Economides, J. Rosenthal, Clinical development success rates for investigational drugs, Nat Biotechnol. 32(1) (2014 Jan) 40-51. https://doi.org/10.1038/nbt.2786.

[3] S. M. Paul, D.S. Mytelka, C.T. Dunwiddie, C.C. Persinger, B.H. Munos, S.R. Linborg, A.L. Schacht, How to improve R&D productivity: the pharmaceutical industry's grand challenge, Nat Rev Drug Discov. 9(3) (2010 Mar) 203-214. https://doi.org/10.1038/nrd3078.

[4] E. Pulkstenis, K. Patra, J. Zhang, A Bayesian paradigm for decision-making in proof-of-concept trials. Journal of Biopharmaceutical Statistics 27(3) (2017) 442-456. https://doi.org/10.1080/10543406.2017.1289947.

[5] C. Chuang-Stein, S. Kirby, J. French, K. Kowalski, S. Marshall, M.K. Smith et al., A Quantitative Approach for Making Go/No-Go Decisions in Drug Development, Drug Information Journal 45(2) (2011) 187-202. https://doi.org/10.1177/009286151104500213.

[6] M.E. Cartwright, S. Cohen, J.C. Fleishaker, S. Madani, J.F. McLeod, B. Musser et al., Proof of Concept: A PhRMA Position Paper with Recommendations for Best Practice, Clinical Pharmacology & Therapeutics 87(3) (2010 Mar) 278-285. https://doi.org/10.1038/clpt.2009.286.

[7] R. Fisch, I. Jones, J. Jones, J. Kerman, G.K. Rosenkranz, H. Schmidli, Bayesian Design of Proof-of-Concept Trials, Ther Innov Regul Sci. 49(1) (2015) 155-162. https://doi.org/10.1177/2168479014533970.

[8] S. Roychoudhury, N. Scheuer, B. Neuenschwander, Beyond p-values: A phase II dual-criterion design with statistical significance and clinical relevance, Clinical Trials 15(5) (2018) 452-461. https://doi.org/10.1177/1740774518770661.

[9] R.L. Lalonde, K.G. Kowalski, M.M. Hutmacher, W. Ewy, D.J. Nicholas, P.A. Milligan et al., Model-based drug development, Clin Pharmacol Ther. 82(1) (2007 Jul) 21-32. https://doi.org/10.1038/sj.clpt.6100235.

[10] P. Frewer, P. Mitchell, C. Watkins, J. Matcham, Decision-making in early clinical drug development, Pharm Stat. 15(3) (2016 May) 255-63. https://doi.org/10.1002/pst.1746.

[11] F. Bretz, J.C. Pinheiro, M. Branson, Combining multiple comparisons and modeling techniques in dose-response studies, Biometrics 61(3) (2005 Sep) 738-748. https://doi.org/10.1111/j.1541-0420.2005.00344.x.

[12] B. Bornkamp, J.C. Pinheiro, F. Bretz, MCPMod: An R Package for the Design and Analysis of Dose-Finding Studies, Journal of Statistical Software 29(7) (2009) 1-23

[13] Pat Mitchell, ASA Biopharmaceutical Section Regulatory-Industry Statistics Workshop. 2017